\documentclass[%
print,
superscriptaddress,
 amsmath,amssymb,
aps,
floatfix,
twocolumn
]{revtex4-2}

\usepackage{graphicx}
\usepackage{dcolumn}
\usepackage{bm}
\usepackage{braket}
\usepackage{verbatim}

\usepackage{color}

\newcommand{\Edit}[1]{\textcolor{blue}{#1}}

\usepackage[normalem]{ulem}

\renewcommand{\sout}[1]{}
\renewcommand{\Edit}[1]{#1}

\begin{document}

\preprint{APS/123-QED}

\title{Synchronization of Coupled Kuramoto Oscillators\\ under Resource Constraints}

\author{Keith A. Kroma-Wiley}
\address{
 Department of Physics \& Astronomy, University of Pennsylvania, Philadelphia, PA 19104 USA
}

\author{Peter J. Mucha}
\address{
 Departments of Mathematics and Applied Physical Sciences, University of North Carolina, Chapel Hill, NC 27599 USA
}

\author{Danielle S. Bassett}
\address{ 
Departments of Physics \& Astronomy, Bioengineering, Electrical \& Systems Engineering, University of Pennsylvania, Philadelphia, PA 19104 USA\\
}
\address{ 
Santa Fe Institute, Santa Fe, NM 87501 USA\\
}



\begin{abstract}
A fundamental understanding of synchronized behavior in multi-agent systems can be acquired by studying analytically tractable Kuramoto models. However, such models typically diverge from many real systems whose dynamics evolve under non-negligible resource constraints. Here we construct a system of coupled Kuramoto oscillators that consume or produce resources as a function of their oscillation frequency. At high coupling, we observe strongly synchronized dynamics, whereas at low coupling we observe independent oscillator dynamics, as expected from standard Kuramoto models. For intermediate coupling, which typically induces a partially synchronized state, we empirically observe that (and theoretically explain why) the system can exist in either (i) a state in which the order parameter oscillates in time, or (ii) a state in which multiple synchronization states are simultaneously stable. Whether (i) or (ii) occurs depends upon whether the oscillators consume or produce resources, respectively. Relevant for systems as varied as coupled neurons and social groups, our study lays important groundwork for future efforts to develop quantitative predictions of synchronized dynamics for systems embedded in environments marked by sparse resources.
\end{abstract}

\maketitle

\Edit{\section{Introduction}}
Since their development in 1975 \cite{Kuramoto1975}, coupled Kuramoto oscillators have been used to model a variety of physical systems \cite{Ermentrout1991,Sompolinsky1990,Kozyreff2001,Filatrella2008} and understand associated behaviors \cite{wiley2006size,Gomez-Gardenes2011,Pazo2009,Hong2011,Acebron2005,Olmi2015}. These applications typically require alterations to the canonical model, including the addition of time-delayed coupling or inertia. However, there is one pervasive discrepancy between the Kuramoto model and real systems that remains relatively unstudied: non-negligible constraints on system resources.

Consequent to thermodynamics, no system can sustain oscillations indefinitely without a supply of some resource. Further, the transition from persistent oscillation to the stationary state is typically not discontinuous as a function of resource supply; it is instead gradual. A mechanical watch, for example, will not suddenly stop ticking as the spring loses its stored potential energy, but will gradually lose seconds over time until the spring has fully unwound. \Edit{Note that we will not consider the case where the watch has fully wound down and hence ceased ticking. The analogy serves only to highlight the change in frequency as a function of some resource in physical systems.} While the dependence on resources may be fairly unimportant during periods of adequate supply, that dependence becomes critical to system behavior when resources become scarce. For example, oxygen deprivation during concussion can perturb the activity of coupled neurons in the human brain, and a scarcity of natural resources can alter predator-prey population cycles in ecology.

\Edit{\section{Model}}

We propose a simple model accounting for resource constraints in a system of Kuramoto oscillators. Oscillators are assumed to have resource-dependent internal velocities, to consume resources as a function of their {net} velocities, and to acquire resources from baths of the resource whose levels are unique to each oscillator. That is, oscillator $i$ is associated with a resource level, $R_i$, which modifies its internal velocity, $\omega_i(R_i)$, and is connected to its own thermodynamic resource bath level, $B_i$. A subscript is included on the function $\omega_i(\cdot)$ because the functional dependence between oscillation rate and resource level may itself differ among oscillators. Unlike \cite{Nicosia2017}, the rate of change of the resource level here is a function of the net velocity and the difference between the current resource level and its bath. Thus, in the most general form, the phases, $\phi_i$, and resource levels, $R_i$, of our system obey 
\begin{eqnarray}
    \dot\phi_i &=& \omega_i(R_i) + \frac{K}{N} \sum_j A_{ij}\sin(\phi_j - \phi_i)\,, \\
    \mathrm{and~~} \dot R_i &=& f_i(B_i - R_i,\dot\phi_i)\,,
\end{eqnarray} 
for $N$ oscillators coupled with strength $K$ according to adjacency matrix elements $A_{ij}$.

We simplify this general model by assuming linear relationships for $\omega_i(R_i) = \nu_i + \mu_i R_i$ and $f_i(B_i - R_i,\dot\phi_i) = \alpha_i + D_i (B_i - R_i) + \beta_i \dot\phi_i$. We also assume oscillators are identical in their functional dependencies (i.e.\  $\omega_i(x) = \omega_j(x)$ and $f_i(x,y) = f_j(x,y)$), and we thus drop associated $i$ indices from the expansion coefficients. \Edit{\sout{To eliminate one additional parameter, we redefine $\dot\phi_i$ to $\dot\phi_i - \nu$. Thus $\beta\dot\phi_i$ becomes $\beta\dot\phi_i - \beta\nu$, and we absorb this second term into our definition of $\alpha$.}}  \Edit{Our dynamical equations then become}
\begin{eqnarray}
    \Edit{\dot\phi_i} &\Edit{=}& \Edit{\nu +\mu R_i + \frac{K}{N}\sum_j A_{ij}\sin(\phi_j - \phi_i)\,,} \\
    \Edit{\mathrm{and~~} \dot R_i} &\Edit{=}& \Edit{\alpha + D(B_i - R_i) + \beta\dot\phi_i\,.}
\end{eqnarray}
\Edit{Now we redefine $\dot\phi_i \rightarrow \dot\phi_i + \nu$ to eliminate the $\nu$ parameter from the phase equation. This process will result in the addition of a term $\beta\nu$ into the resource equation, which may be eliminated by taking $\alpha\rightarrow \alpha-\beta\nu$.} 

\Edit{Further, although the frequency variation is due to the variation of resource levels, it will be simpler to work directly with the frequencies. To this end, we take $\mu R_i \rightarrow \omega_i$, which in turn means $\dot R_i \rightarrow \mu^{-1}\dot\omega_i$. We may eliminate this residual $\mu$ parameter by taking $\alpha\rightarrow\mu^{-1}\alpha$ and $\beta\rightarrow\mu^{-1}\beta$. We will also define $W_i \equiv \mu^{-1}B_i$. We may also absorb $\alpha$ into the definition of $W_i$ by taking $W_i + \alpha/D \rightarrow W_i$. After these steps, our dynamical equations are}
\begin{eqnarray}
    \Edit{\dot\phi_i} &\Edit{=}& \Edit{\omega_i + \frac{K}{N}\sum_j A_{ij}\sin(\phi_j - \phi_i)\,,} \\
    \Edit{\mathrm{and~~} \dot \omega_i} &\Edit{=}& \Edit{D(W_i - \omega_i) + \beta\dot\phi_i\,.}
\end{eqnarray}
\Edit{By this process, we have reduced our system to the minimal number of unique parameters, up to a redefinition of the timescale, which could technically eliminate one additional parameter.} 

\Edit{To conform to common notation, we make one final transformation of the system by plugging the $\dot\phi_i$ equation into the frequency equation, making the Kuramoto contributions to the frequency dynamics explicit. This also introduces a $\beta\omega_i$ term into the frequency equation, but this may be absorbed into the definitions of $D$ and $W_i$ by taking $D' = D-\beta$ and $W_i' = W_i/(D-\beta)$. Thus, our final equations are}
\begin{eqnarray}
    \Edit{\dot\phi_i} &\Edit{=}& \Edit{\omega_i + \frac{K}{N}\sum_j A_{ij}\sin(\phi_j - \phi_i)\,,} \\
    \Edit{\mathrm{and~~} \dot \omega_i} &\Edit{=}& \Edit{D'(W_i' - \omega_i) + \beta\frac{K}{N}\sum_j A_{ij}\sin(\phi_j - \phi_i)\,.}
\end{eqnarray}
\Edit{\sout{In words, $\mu$ defines the sensitivity of internal velocity to resource level, $\alpha$ defines the intrinsic production (if positive) or consumption (if negative) of resource by an oscillator,}} In words, $D$ defines the ``diffusion'' rate of the frequency toward $W'_i$, the equilibrium frequency when $\beta=0$, and $\beta$ defines the amount of resource consumed (if negative) or produced (if positive) per oscillation. \Edit{For our main results, we will take the connectivity to be all-to-all, so that $A_{ij} = 1$ for all $i\neq j$. We will comment briefly at the end of this paper on the behavior of another topology, but non-trivial topologies are pragmatically much more subject to finite-size effects since simulation runtime increases as $N^2$ (at fixed coupling density) when matrix calculations become necessary. In the all-to-all case, we avoid this difficulty because the mean-field approach becomes exactly, rather than approximately, correct.}

\Edit{In the form of Equations 7 and 8, it becomes clear that this model is essentially a modification of the frequency adaptation model studied by Taylor \emph{et al.}~\cite{Taylor2010}, which is itself a modification of Kuramoto models with inertia \cite{Acebron1998}. However, their model considered only $\beta>0$, included a noise term in the resource dynamics, and assumed that the potential was the same for all oscillators, equivalent in this model to assuming all $W_i$ are equal. This means that the only force impeding synchronization in that model was the noise term, since the oscillators were identical otherwise. This leads to quite significant differences between the model they considered and the one studied here. Later work in this area tended to focus on adaptive frequencies \cite{Skardal2014} or somewhat simpler models of frequency adaptation \cite{Ott2017}, chosen to be more easily amenable to the Ott-Antonsen ansatz \cite{Ott2008}.} 

\begin{figure}
    \includegraphics{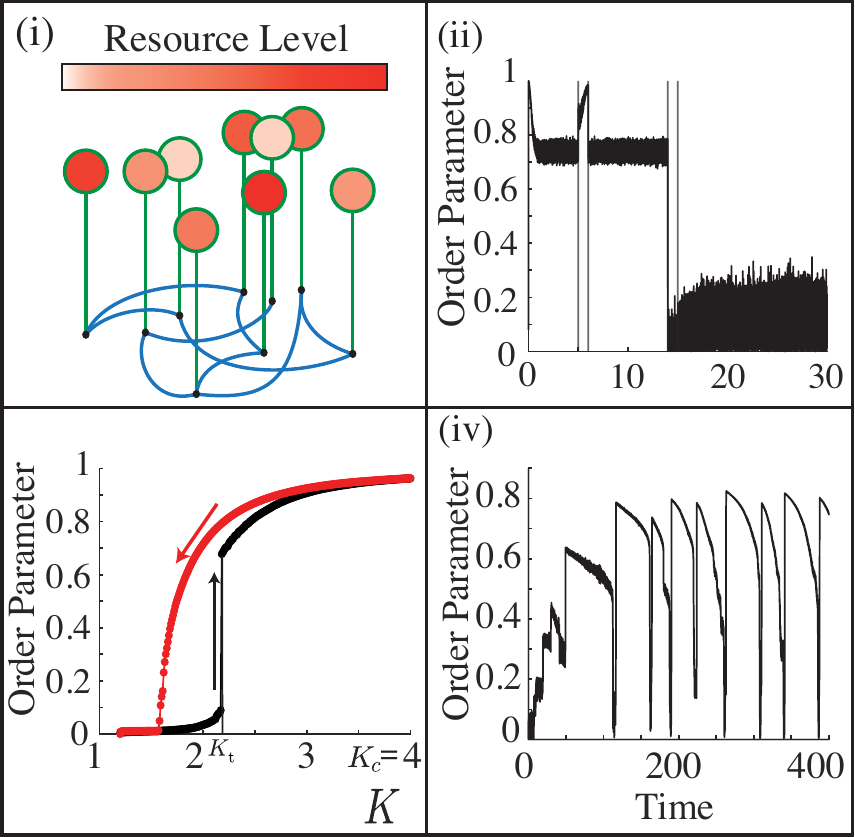}
    \caption{\textbf{Overview of System Behavior.} \emph{(i)} Schematic of system configuration of oscillators (black points), resource baths (red circles), diffusive connections (green lines), and phase-coupled interactions (blue lines). \emph{(ii)} Time series of the order parameter in the bistable case. During time 5--6, a perturbation forces the system to synchronize. During time 15--16, a perturbation forces the system to desynchronize. \emph{(iii)} The temporal mean and standard deviation of the order parameter as a function of the coupling strength, shown both for increasing (black) and then decreasing (red) the coupling strength. We observe swift synchronization when increasing the coupling strength, and bistability is evidenced by the different synchronization levels observed for the same value of the coupling strength. \emph{(iv)} A sample time series of the oscillating order parameter regime. Parameters in (ii) and (iii) are $N=6.5\times 10^6$, $\beta=3$, $D=5$, $\mu_W = 0$, $\sigma_W = 1$, $\Lambda = 10$, and in (ii) $K = 2$. In (iv), $N=2\times 10^4$, $K = 1.5$, $\beta=-2$, $D=1$, $\mu_W=0$, $\sigma_W=1$, $\Lambda=100$. Note that $\mu_W$ and $\sigma_W$ are the mean and standard deviation of the Gaussian distributed bath frequencies $W_i$.}
    \label{Fig:MainFigure}
\end{figure}

\Edit{\section{Simulation}}

To probe the behavior of this model, we study a system of $N = 6.5\times10^6$ oscillators with normally distributed \Edit{\sout{$B_i$} $W_i$}. \Edit{\sout{Oscillators are coupled on an Erd\H{o}s-R\'enyi network with edge probability 0.2, which is large enough to ensure mean field theory can provide a reasonable approximation of the system's behavior.}} For analytic simplicity, model parameters are chosen so that the phase timescale is much faster than the resource timescale. \Edit{In particular, we scale $W_i$ and $K$ by some large scalar $\Lambda$, and we scale $D$ and $\beta$ by a factor of $\Lambda^{-1}$. As a result of the scaling of $\Omega_i$ and the diffusive nature of the resource dynamics, $\omega_i$ will also be $\mathcal{O}(\Lambda)$ after some equilibration time. Thus, $\dot\phi$ will be scaled by a factor of $\Lambda$, while $\dot\omega$ will be left unchanged, thereby separating the timescales, which is the very thing it was required to do.} Varying $K$ and $\beta$, we evaluate synchronization using the order parameter $r$ defined by $r e^{i\psi} = \frac{1}{N}\sum_j e^{i\phi_j}$ (Fig.~\ref{Fig:MainFigure}). In this sample instantiation, we observe four primary behaviors: (i) a fixed point at asynchrony for small $K$, (ii) a fixed point at synchrony for large $K$, (iii) an oscillating order parameter if $K$ is intermediate and $\beta < 0$, and (iv) bistability if $K$ is intermediate and $\beta > 0$. No other significantly different behaviors were observed for other parameter choices, although some may exist given the size of the parameter space. 

\Edit{As an aside, one may wonder why we have not considered the opposite limit, where the resource dynamics are \emph{faster} than the phase dynamics, i.e.\ the small $\Lambda$ limit. The reason is that, in the limit of small $\Lambda$, the dynamics become standard Kuramoto dynamics. To see why this is so, note that if the phase timescale is infinitely slower than the resource timescale, then the frequency $\omega_i$ will perpetually be at its equilibrium level. Thus, we can take $\dot\omega_i=0$, solve for $\omega_i$, and plug this expression directly into the phase equation. Referring to the form of Equation 6, we may set $\dot\omega_i = 0$ and solve for $\omega_i$ to obtain}
\begin{equation}
    \Edit{\omega_i = W_i + \frac{\beta}{D}\dot\phi_i}\,.
\end{equation}
\Edit{This expression may in turn be plugged into Equation 5 and solved for $\dot\phi_i$ to obtain}
\begin{equation}
    \Edit{\dot\phi_i = \bigg(1 - \frac{\beta}{D}\bigg)^{-1}\bigg[W_i + \frac{K}{N}\sum_j A_{ij}\sin(\phi_j - \phi_i)\bigg]\,.}
\end{equation}
\Edit{And so we see that in the small $\Lambda$ limit, we have standard Kuramoto dynamics with an altered timescale, with natural frequencies distributed according to $W_i$. Figure 3 in the Supplementary Material (SM) \cite{SM} demonstrates that the seemingly first-order transition observed in Fig.~1.iii becomes second-order as $\Lambda \to 0$.}

\Edit{\section{Analysis}}

We first consider the possible behaviors of a single oscillator interacting with a supposed synchronized group, and then we turn to group-level dynamics. An oscillator can be considered a two-state system (Fig.~2.i). In the first state, \emph{S1}, the oscillator turns with a net velocity equal to the average velocity of the group ($\dot\phi_i = \Omega$); in the second state, \emph{S2}, the oscillator turns with a net velocity equal, on average, to its internal velocity ($\dot\phi_i = \omega_i$\Edit{\sout{ $=  \mu R_i$}, note that this is the assumption that breaks down when $\Lambda \ll 1$}). Because the equilibrium \Edit{\sout{resource level} internal velocity} is defined in part by the net velocity, \Edit{\sout{and because the internal velocity is defined by resource level,}} each of the states \emph{S1} and \emph{S2} will have corresponding equilibrium internal velocities $\omega_i($\emph{S1}$)$ and $\omega_i($\emph{S2}$)$ that the oscillator will approach over time. Further, an effective coupling strength is defined in mean field theory, given the size of the current synchronized group and the coupling strength. This effective coupling defines an envelope in frequency space whose boundaries are $\Omega \pm Kr$\Edit{\sout{$/N$}} (Fig.~2.ii). See Sections 1 and 2 of the SM for a derivation of this envelope. If an oscillator's internal velocity $\omega_i$ places it within the envelope, it will be captured by (i.e.~synchronize with) the group, and if $\omega_i$ places it outside of the envelope, it will remain free (i.e.~unsynchronized).

The possible dynamics of a single oscillator can thus be separated into 4 types. A Type 1 (`Stably Free') oscillator is one for which both $\omega_i($\emph{S1}$)$ and $\omega_i($\emph{S2}$)$ are outside of the envelope, so that the oscillator will always eventually transition to state \emph{S2}. A Type 2 (`Stably Captured') oscillator is one for which both $\omega_i($\emph{S1}$)$ and $\omega_i($\emph{S2}$)$ are inside of the envelope. A Type 3 (`Transitory') oscillator has $\omega_i($\emph{S1}$)$ outside the envelope and $\omega_i($\emph{S2}$)$ inside the envelope and will continually alternate between capture and escape. A Type 4 (`Bistable') oscillator has $\omega_i($\emph{S1}$)$ inside the envelope and $\omega_i($\emph{S2}$)$ outside the envelope, so both states \emph{S1} and \emph{S2} will be stable. Note that these four behaviors do not depend on any of the simplifications made to the general model. Further, note that the four different group behaviors are not in one-to-one correspondence with these four single-oscillator behaviors, as will be elaborated below. Finally, in the context of our simplified model, we can show that if \Edit{$\beta > 0$ \sout{and $\mu$ have the same sign}}, then Type 4 (`Bistable') oscillators may exist; whereas if \Edit{$\beta>0$ \sout{and $\mu$ have opposite signs}}, then Type 3 (`Transitory') oscillators may exist (see Section 5 of the SM). Thus, since the \Edit{\sout{choices} choice of $\beta$ is \sout{and $\mu$ are}} system-wide, only one of Type 3 and Type 4 oscillators may exist.

\begin{figure}
    \includegraphics{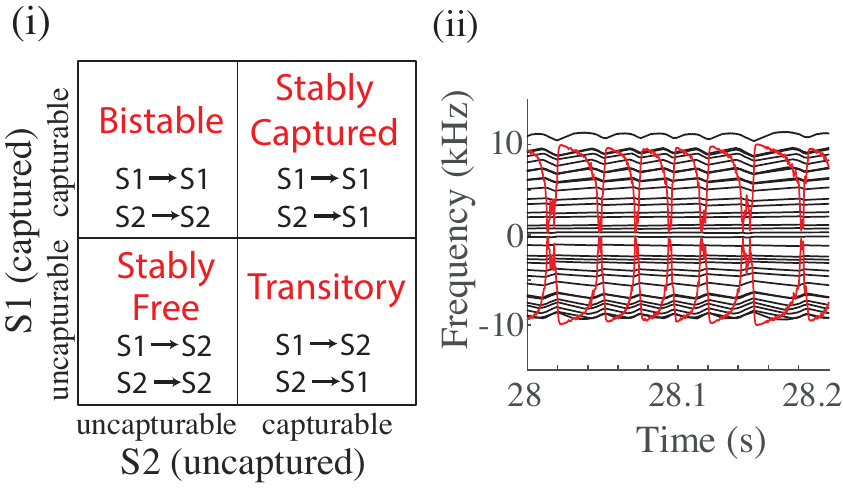}
    \caption{\textbf{Available State Transitions.} \emph{(i)} The four possible states of a given oscillator subject to its present effective coupling strength to the rest of the group. \emph{(ii)} The envelope in which an oscillator can be captured. Here the effective coupling strength (${Kr}$\Edit{\sout{$/{N}$}}; red) is overlaid on the distance of each oscillator's internal velocity from the group velocity, defined as the average of all internal velocities, and scaled by the degree of that oscillator (black).}
    \label{Fig:FlowLines}
\end{figure}

Having defined the behavior of a single oscillator interacting with a dominant synchronized group, we now turn to the question of group behaviors. Such group behaviors can be understood by considering the effects of dynamics on the \Edit{\sout{available resources} internal velocities}. A Type 1 (`Stably Free') oscillator's \Edit{\sout{resource level} internal velocity} will equilibrate to that characteristic of desynchronization. A Type 2 (`Stably Captured') oscillator's \Edit{\sout{resource level} internal velocity} will equilibrate to that characteristic of synchronization. The \Edit{\sout{resource level} internal velocity} of a Type 3 (`Transitory') oscillator will oscillate back and forth between levels characteristic of synchronization and desynchronization, and as a result will pile up at the boundaries of the synchronization envelope. By symmetry of the envelope, Type 3 oscillators will pile up both above and below the \Edit{\sout{mean resource level} group frequency}, and so \Edit{\sout{both the resource distribution and in turn}} the internal velocity distribution of Type 3 oscillators will become bimodal over time. For Type 4 oscillators, their distribution will be smooth over \Edit{\sout{resource levels} internal velocities}, but will depend on the history of the system.

\begin{figure*}
    \includegraphics[width=0.85 \textwidth]{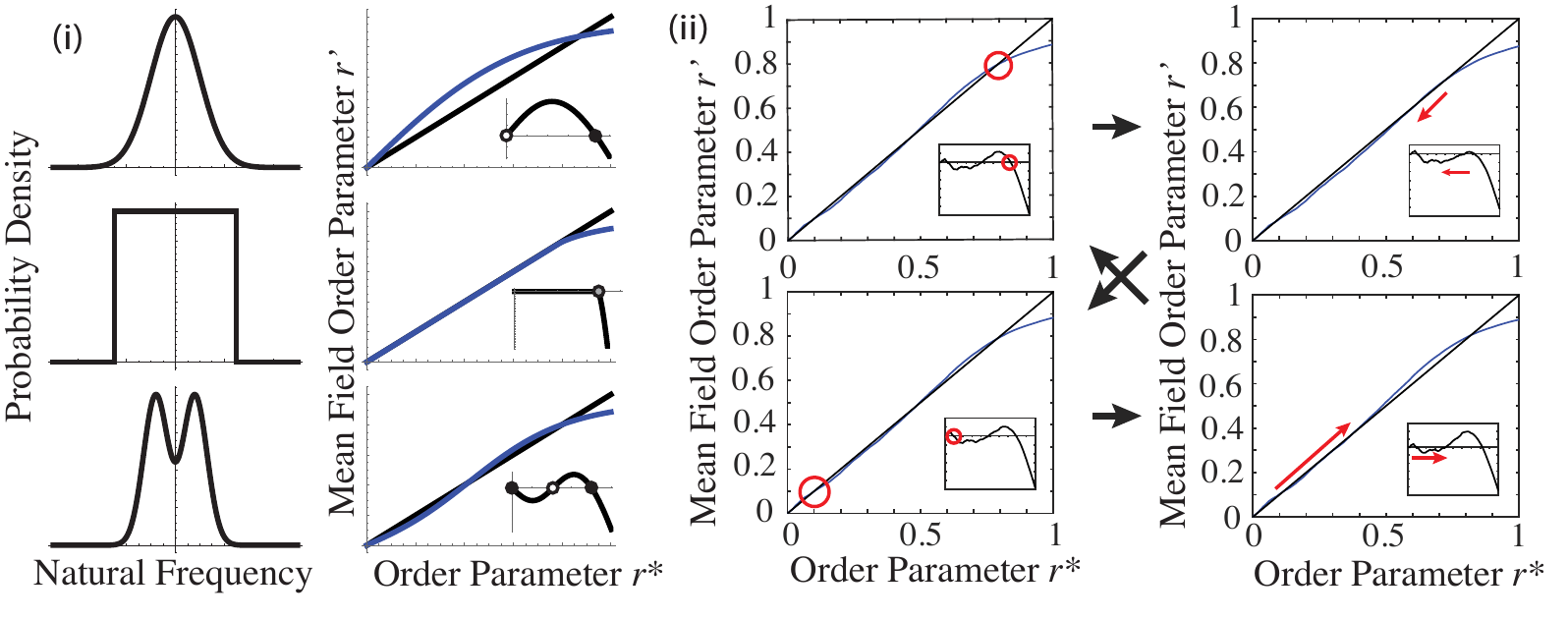}
    \caption{\textbf{Evolving Order of Type 3 `Transitory' oscillators.} \emph{(i)} \emph{Left} Example natural frequency distributions. \emph{Right} Graphical solutions to the mean field system for the adjacent natural frequency distributions. The present order parameter of the system is marked by $r^{*}$. The mean-field order parameter with effective coupling strength $K_{\text{eff}} = Kr^{*}\Edit{\sout{\braket{d}/N}}$ is marked by $r'$. The blue curve gives $r'$ as a function of $r^{*}$, while the black curve is the line $r' = r^{*}$. Points where the blue and black curves intersect are predicted steady-states of the order parameter for the system. The inset figures give the difference between the blue and black curves. Open circles in the inset figures represent unstable equilibria; black circles represent stable equilibria; and the grey circle represents metastable equilibrium. The top two subfigures give the solution for a Gaussian distribution of natural frequencies, the middle two for a uniform distribution, and the bottom two for a bimodal internal frequency distribution. \emph{(ii)} The graphical solution at four different moments of the time evolution of the oscillating order parameter system. Red circles indicate the present order parameter of the system. Red arrows indicate that the system's order parameter is quickly changing in that direction. Black arrows indicate the temporal sequence of the four states. States were chosen such that the top left was a moment of stable synchronization, top right was a moment of desynchronization, bottom left was a moment of stable disorder, and bottom right was a moment of resynchronization.}
    \label{Fig:OrderEvolution}
\end{figure*}

In order to understand the transitions observed in the order parameter, we consider graphical solutions to a mean field Kuramoto system described by
\begin{equation}
    r = \Edit{\sout{\frac{Kr\braket{d}}{N}}}Kr\int_{-\frac{\pi}{2}}^{\frac{\pi}{2}} d\omega\, g\bigg(\Edit{\sout{\frac{Kr\braket{d}}{N}}}Kr\sin{\omega}\bigg)\cos^2(\omega)\,,
    \label{Eq:Order}
\end{equation}
where $\omega$ is the internal velocity\Edit{\sout{, $\braket{d}$ is the average degree,}} and $g(x)$ is related to the probability distribution of internal velocities (see Fig.~\ref{Fig:OrderEvolution} and Sections 1 and 2 of the SM for derivation). Solutions at a chosen moment provide order parameters that would be stable in the mean field description of a system of Kuramoto oscillators with natural velocities equal to their value at that same chosen moment. In fact, the natural velocities of the oscillators drift over time, but because the phase timescale is much faster than the resource timescale, the system will equilibrate its phase behavior before resource levels change appreciably, and therefore this equation is sufficient.

To gain further intuition, suppose the system is presently in some synchronization state with order parameter $r^{*}$ (Fig.~\ref{Fig:OrderEvolution}). The right side of Eq.~\ref{Eq:Order} then gives the equilibrium order parameter of the system if each oscillator interacted with a single mean field oscillator with coupling strength $Kr^{*}$\Edit{\sout{$\braket{d}/N$}}. Call this order parameter $r'$. If $r' > r^{*}$, we expect order in the system to increase; if $r' < r^{*}$, we expect order to decrease. This simple argument captures the stability properties of many Kuramoto systems. For example, it correctly predicts the stability properties of Gaussian, \Edit{\sout{unimodal}uniform}, and (weakly) bimodal natural frequency distributions (Fig.~\ref{Fig:OrderEvolution}.i). 

Now we have all of the tools to understand the time evolution of the oscillatory order that we observed. Suppose we let a system of oscillators run for some period of time, and at a given moment we find the order parameter to be large. For a given present value of the order parameter, some of our oscillators will be Type 1, some Type 2, and some Type 3. The resource levels of Type 1 and Type 2 oscillators will be spread over some range, but the Type 3 oscillators will drive a bimodal resource distribution, generating bistability in the system (Figure 3.ii, top left). For the system to be in a high order state, it must be that many of the Type 3 oscillators are currently synchronized inside of the envelope. Thus, these oscillators will gradually move outward toward the boundary of the synchronization envelope. At the group level, this process has the effect of pushing the upper bump of the blue curve downward. Eventually, a saddle-node bifurcation occurs and the high-order solution vanishes. Thus, the system must transition to the low-order state, causing the synchronization envelope to shrink dramatically. Consequently, all of the Type 3 oscillators, no matter their present state, will desynchronize (Fig.~\ref{Fig:OrderEvolution}.ii, top right). 

As a result of now being outside of the envelope, the resource levels of all Type 3 oscillators will now move quickly toward the mean level, clustering closely together in resource space. The high-order solution will almost immediately reappear as the Type 3 oscillators begin to cluster together (Fig.~\ref{Fig:OrderEvolution}.ii, bottom left, reflecting a moment of stable disorder). However, since the low-order solution is stable, the system will persist in the low-order state until the Type 3 oscillators cluster sufficiently that the low-order state becomes unstable. For reasons discussed in Section 3 of the the SM, in a finite-sized system, this will be the moment when the ``energy barrier'' between the low- and high-order states becomes sufficiently small that the system can randomly transition to the high-order state (Fig.~\ref{Fig:OrderEvolution}.ii, bottom right). In an infinite-sized system, the details of this transition are unknown to the present authors. Once such a transition occurs, we find the system again in a state of high order (Fig.~\ref{Fig:OrderEvolution}.ii, top left), and the cycle repeats.

\begin{figure}
    \includegraphics[scale=0.8]{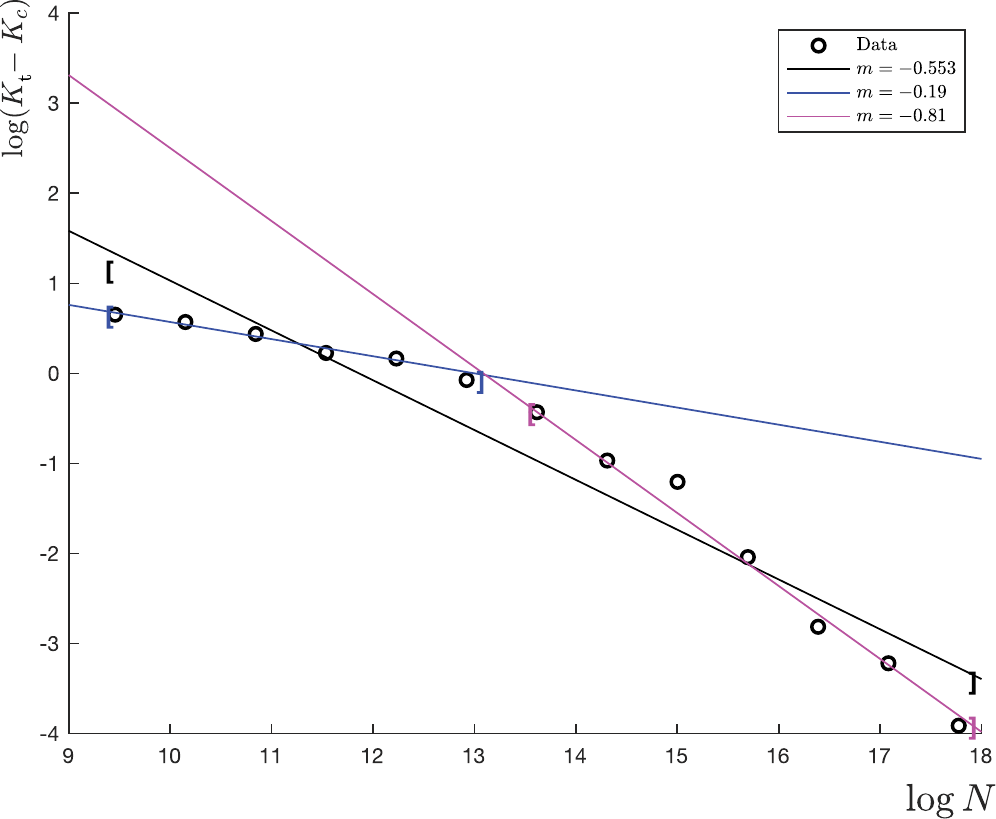}
    \caption{\textbf{Finite-Size Scaling of $\mathbf{K_t}$} Here the system is simulated with the same parameters as were used in Figure 1.iii, except that the system size $N$ is varied over a wide range. The logarithm of the system size is plotted against the logarithm of the difference between $K_t$ and $K_c$. Note that as system size increases, the difference between $K_t$ and $K_c$ becomes as small as desired. However, the size dependence is not simple. It goes as $N^{-0.2}$ for smaller system sizes, but appears to tend toward $N^{-1}$ as system size increases.}
    \label{Fig:FiniteSizeScaling}
\end{figure}

Now that we understand group behaviors for Type 3 oscillators, we turn to Type 4 oscillators. Recall that an oscillator's classification depends upon the present effective coupling strength $Kr$. With zero coupling, the system will be made up entirely of Type 1 oscillators. If we then increase the coupling, some of the oscillators will become Type 4 oscillators, which being previously unsynchronized, remain unsynchronized. This point in the system's dynamics is precisely a subcritical bifurcation, since if we moved all of the Type 4 oscillators inside of the envelope, they would remain there and form a stable synchronized group. As we continue to increase the coupling strength, eventually some of the Type 4 oscillators will become Type 2 oscillators, and a small synchronized group of Type 2 oscillators will appear. Upon changing from Type 4 to Type 2 oscillators, the internal velocities will transition from $\omega_i($\emph{S1}$)$ to $\omega_i($\emph{S2}$)$, causing the order parameter of the system to continually increase, even without changes in the coupling strength. The additional growth will capture more oscillators, initiating a cascade of synchronization. Thus, the system will experience a transition up to the synchronized phase which appears to be first-order in the coupling strength, commonly referred to as explosive synchronization \cite{dsouza2019explosive}. In the limit of infinite system size, this transition will occur at the critical coupling strength $K_c$ of the unsynchronized natural frequency distribution. In Figure 1.iii, this occurs roughly at $K=4$. However, we see that in fact the transition coupling strength $K_t$ is actually only a bit more than 2 in that plot. However, as the system size increases, $K_t$ does trend toward the expected value of $K_c$ very slowly, as seen in Figure \ref{Fig:FiniteSizeScaling}.

If we then turn the coupling strength down slowly, the Type 2 oscillators will again become Type 4 oscillators as the synchronization envelope narrows, but high order will persist. Eventually, some of the Type 4 oscillators will become Type 1 oscillators, and so move outside of the envelope of synchronization. However, unlike the case of synchronization, all unsynchronized oscillators are equivalent in their effect on the order parameter. Thus, the transition from $\omega_i($\emph{S2}$)$ to $\omega_i($\emph{S1}$)$ will not cause additional change in the order parameter. Consequently, the transition back to disorder will be continuous in the coupling strength; in other words, a second order phase transition. This mixture of first- and second-order behavior suggests that the normal form of the bifurcation cannot be a simple one-dimensional form, as can most Kuramoto model modifications.

\Edit{\section{Hierarchically Modular Topology}}
\Edit{Following the work of Arenas \emph{et al.}~\cite{Arenas2006}, we also investigate the behavior of the model dynamics occurring atop a hierarchically modular topology in which small modules are embedded within larger modules. Specifically, suppose there are three hierarchical levels in the network. At the lowest level, there are many small communities which are very densely connected. At the intermediate level, these small communities are connected together more weakly. Finally, at the highest level, the nodes are connected the weakest.}  

\Edit{We aimed to see whether we observed, as Arenas \emph{et al.}\  did, that the order parameter dynamics reflected the topology of the network. In particular, they observed that in such a hierarchically modular network, the order parameter changed in a somewhat stepwise fashion as the strength of the coupling constant grew, with the steps corresponding to synchronization of different levels of the hierarchy. As the coupling increases, first a critical point will be reached where the lowest-level communities synchronize. Then, as the coupling increases further, the next levels of the hierarchy synchronize together, and finally at some point the whole network will synchronize. Thus, by observing just the global dynamics of the order parameter, one can acquire an understanding of some of the details of the network structure.}

\begin{figure}
    \includegraphics[scale=0.8]{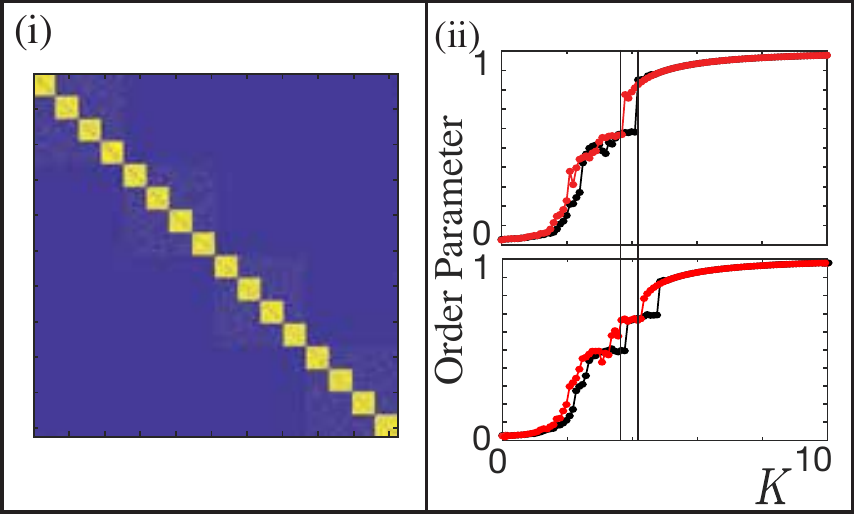}
    \caption{\textbf{Bistable Transition in Hierarchical Network.} (i) Here we plot the adjacency matrix for the hierarchical topology considered. There are 1024 nodes, and each node has 72 edges. Of these, 60 are distributed among its closest hierarchical group, 10 are distributed at the next hierarchical level, and 2 at the remaining level. In (ii) we show two plots of the phase transition of the order parameter as a function of the coupling strength for two different random instantiations of the model with the same parameters. As seen by Arenas \emph{et al.}~\cite{Arenas2006}, the order in the system shows clear plateaus as $K$ increases. However, in our case it appears that if the plateaus are too close together, as seems to be the case between the second and third in the upper plot, the transition jumps immediately from the first to the third plateau. We also see the first-order transition between the plateaus that we have come to expect from the bistable regime. The stark different between the two curves also serves to highlight how far we are from the thermodynamic limit. Vertical lines are drawn in for visual comparison.}
    \label{Fig:HierarchicalFigure}
\end{figure}

\Edit{To assess the dependence of the order parameter on the network hierarchy, we follow the hierarchical framework of Ref.~\cite{Arenas2006}, but with a larger number of nodes to improve the numerical stability of our present results. We construct a network with 1024 nodes with 16 communities of 64 nodes each which are connected internally with degree $z_1$. We then group these communities into sets of four and let each node have $z_2$ connections with nodes from the other three communities in this coarser hierarchical level. Finally, we let each node have $z_3$ connections with nodes that are in neither of the first or second hierarchical levels. Thus, each node will have degree $z_1 + z_2 + z_3$. An example adjacency matrix with $z_1 = 60$, $z_2 = 10$, $z_3 = 2$ is presented in Figure \ref{Fig:HierarchicalFigure}.i.}

\Edit{For the bistable configuration of our resource constrained model, we hypothesized that the steps between order parameter plateaus would be sharper than those observed by Arenas \emph{et al.}, since our transition from low to high order is first-order, whereas their transition was second-order. However, due to the synchronization cascade one could alternatively hypothesize that once the lowest hierarchical level synchronizes, the effective coupling would gradually become strong enough that each subsequent hierarchical level synchronizes in turn; in such a case, we would observe a single step in the order parameter, thus obfuscating any reflection of the network structure in the dynamics. In reality, it is difficult to validate either hypothesis although clear structure exists in the order parameter curves presented in Figure \ref{Fig:HierarchicalFigure}.ii.}

\Edit{As Arenas \emph{et al.}\ observed, we expect three plateaus of the order parameter, corresponding to each of the three hierarchical levels. The width of some of these plateaus in $K$-space should differ depending on the exact choices of $z_1$, $z_2$, and $z_3$. In our case, we observe both behaviors hypothesized above; that is, apparently discontinuous jumps between plateaus, as well as skipping plateaus altogether. In the lower plot of Figure \ref{Fig:HierarchicalFigure}.ii we see 3 distinct plateaus. The black curve is the upward-going transition and we see discontinuous jumps between each of the plateaus. In the downward-going red curve, the transitions are smoother, as expected. This behavior validates the first hypothesis. However, for another run of the model with the same parameters, simply allocated differently due to the random nature of edge assignment, we observed that the system never moved to the second plateau. Rather, it jumped directly from the first plateau to the third. This behavior is seen in the upper plot of Figure \ref{Fig:HierarchicalFigure}.ii.}

\Edit{Collectively, these findings suggest that non-trivial network topology can be directly reflected in order parameter dynamics. Yet, we caution that several limitations of our findings should be considered. Our system size is very small due to the long runtime of matrix operations with large adjacency matrices. Thus, we cannot claim to be in an infinite-size limit, and so the mean-field dynamics we discussed above do not map nicely onto this example case.}

\Edit{\section{Discussion}}

The behavior of this Kuramoto model can clarify dynamics observed in many physical and biological systems existing under resource constraints. For example, transient oscillatory synchrony has been observed to be important for efficient routing of information \cite{Palmigiano2017} in neural systems. Consider also alcohol consumption in a group of acquaintances; each individual will have a preferred frequency of consumption, but their actual frequency will be modulated by interaction with the group. Each individual also accrues the resource of craving at a different rate based on environmental stimuli such as advertisements, experiences, and inter-personal communication \cite{seo2014neurobiology}. Consumption can either increase or decrease craving; for some, satiating the urge, while for others, strengthening it. Oscillations in this context are reminiscent of binge drinking behaviors, in which an individual is usually satisfied to drink at a group frequency, but eventually their craving grows until they break off and drink at a higher-than-average frequency for a brief period, after which they settle back into the group frequency. Future studies could exercise the model to study such resource-constrained biological, physical, social, and technological systems, and to better understand their associated behaviors.

\Edit{One may reasonably wonder why we have relied only on relatively intuitive arguments throughout this paper, instead of making use of the Ott-Antonsen ansatz \cite{Ott2017}, which is known to greatly reduce the dimensionality of Kuramoto systems. In fact, we do use this ansatz in Section 6 of the SM, where we also use a two-timing approach to simplify the dynamics further, taking advantage of the separation between the phase and diffusion timescales. This approach results in 3 coupled differential equations: two defining the phase dynamics on the fast and slow timescales, and one describing the resource dynamics on the slow timescale. However, we did not gain further insight, even with these differential equations in hand. In \cite{Ott2017}, Ott and Antonsen used their ansatz to study a different model of frequency adaptation, but the application there was simplified by the fact that each oscillator's frequency adaptation dynamics depended linearly on the difference between the current velocity $\omega_i$ and the group velocity $\Omega$. That is, their dynamics were of the form
\begin{eqnarray}
    \dot\phi_i &=& \omega_i + \frac{K}{N} \sum_j A_{ij}\sin(\phi_j - \phi_i)\,, \\
    \mathrm{and~~} \dot \omega_i &=& -\gamma(r)[\omega_i - W_i] - \nu(r)[\omega_i - \Omega]\,.
\end{eqnarray}
where $\nu$ and $\gamma$ are arbitrary functions of only the order parameter of the system.
The closest approach to this form for our model, averaging out the phase behavior in the resource dynamics (i.e.\ replacing the sine term by its average over a period), is
\begin{eqnarray}
    \dot\phi_i &=& \omega_i + \frac{K}{N} \sum_j A_{ij}\sin(\phi_j - \phi_i)\,, \\
    \mathrm{and~~} \dot \omega_i &=& -D[\omega_i - W_i] - \beta f(r,\omega_i)[\omega_i - \Omega]\,
\end{eqnarray}
where
\begin{equation}
    f(r,\omega_i) = \left\{ 
        \begin{array}{ll}
            1 & \quad |\omega_i - \Omega| < Kr \\
            1 - \sqrt{1 - (\frac{Kr}{\omega_i - \Omega})^2} & \quad |\omega_i - \Omega| > Kr\,.
        \end{array}
    \right.
\end{equation}
It is clear that the nonlinear dependence on $\omega_i$ in the second term in $f(r,\omega_i)$ is unaccounted for in the model considered by Ott and Antonsen. We consider this nonlinearity to be important because it arises from the natural assumption that resource consumption is frequency-dependent.}

\Edit{Nevertheless, putting the model into this form does highlight another way of understanding some of the results we observe. In particular, recall that the phase transition in the case of bistability appears to be discontinuous as $K$ increases, but is continuous as K decreases. We can now understand this asymmetry in the following way. The resource dynamics for captured oscillators (i.e.\ oscillators for which $|\omega_i - \Omega| < Kr$) possess no dependence at all on the order parameter. Thus, as the coupling decreases, oscillators will gradually fall away from the edges of the synchronized group, but as long as an oscillator remains captured, its frequency will remain unchanged. Further, unsynchronized oscillators, in the thermodynamic limit, exert no pull on the synchronized oscillators, and so the phase transition will necessarily be exactly identical to that of the standard Kuramoto model. However, in the case where $K$ is increasing and oscillators are moving from uncaptured to captured, the fact that the uncaptured resource level is a function of the order parameter precludes the use of the same argument; thus, a discontinuous transition becomes possible.}

Many open questions remain in this system. The social context above motivates consideration of mixed populations of $\beta>0$ and $\beta<0$ oscillators. In addition, the nature of the connection topology is an important modulator of observed dynamics in standard Kuramoto systems, and we expect the same to be true here. One particularly interesting topological modification would be to connect the baths in the resource level, encoding a multilayer network structure \cite{bianconi2018multilayer} that allows individual oscillators to interact with a shared resource environment. This would cause not just the second term of the resource dynamics to be nonlinear, but also the first term. Although resource-constrained Kuramoto systems have been relatively little-considered up to this point, we hope that the interesting behavior observed in just the simplest version of this model motivates further work moving forward.\\

\noindent \textbf{Acknowledgements.} We thank Lia Papadopoulos, Zhixin Lu, and Erfan Nozari for providing valuable feedback. This work was primarily supported by the Army Research Office (W911NF-18-1-0244), the Paul G. Allen Foundation, and the NSF through the University of Pennsylvania Materials Research Science and Engineering Center (MRSEC) DMR-1720530. PJM also acknowledges support from the James S. McDonnell Foundation 21st Century Science Initiative - Complex Systems Scholar Award grant \#220020315 and the NSF (ECCS-161076). The content is solely the responsibility of the authors and should not be interpreted as representing the official views or policies, either expressed or implied, of any of the funding agencies. The U.S Government is authorized to reproduce and distribute reprints for Government purposes notwithstanding any copyright notation herein.

\bibliography{main}
\bibliographystyle{apsrev4-2}
\frenchspacing

\end{document}